\documentclass[twocolumn,english,aps,prd,nofootinbib]{revtex4-1}
\usepackage{CJK}

\usepackage{amsfonts}
\usepackage{amsmath}
\usepackage{hyperref}
\usepackage{amssymb}
\usepackage{xcolor}
\usepackage{relsize}
\usepackage[bottom]{footmisc}

\usepackage{tikz}
\usetikzlibrary{calc} 

\newcommand{\be}{\begin{equation}}
\newcommand{\ee}{\end{equation}}
\newcommand{\bea}{\begin{equation} \begin{aligned}}
\newcommand{\eea}{\end{aligned} \end{equation} }
\newcommand{\bi}{\begin{itemize}}
\newcommand{\ei}{\end{itemize}}

\usepackage{xspace}

\newcommand{\la}{\lambda}

\newcommand{\lp}{\left(}
\newcommand{\rp}{\right)}

\usepackage{color}
\usepackage{graphicx}
\usepackage[space]{grffile}
\usepackage{verbatim}
\usepackage{amsmath}
\usepackage{amssymb}
\usepackage{wasysym}
\usepackage[caption=false]{subfig}
\usepackage{url}
\usepackage{bbold}
\usepackage{slashed}
\usepackage{epstopdf}
\usepackage{braket}
\usepackage{float}

\DeclareRobustCommand{\Sec}[1]{Sec.~\ref{#1}}

\DeclareRobustCommand{\Fig}[1]{Fig.~\ref{#1}}

\DeclareRobustCommand{\Eq}[1]{Eq.~(\ref{#1})}
\DeclareRobustCommand{\Eqs}[2]{Eqs.~(\ref{#1}) and (\ref{#2})}
\DeclareRobustCommand{\Ref}[1]{Ref.~\cite{#1}}

\DeclareMathAlphabet\mathbfcal{OMS}{cmsy}{b}{n}

\newcommand{\boundellipse}[3]
{[black,fill=blue!30] (#1) ellipse (#2 and #3)
}
\newcommand{\boundellipseW}[3]
{[white,fill=white] (#1) ellipse (#2 and #3)
}

\begin{document}

\title{Solving the quantum dimer and six vertex models one electric field line at a time}

\author{Jonah Herzog-Arbeitman$^{1,2}$}
\author{Sebastian Mantilla$^{2}$}
\author{Inti Sodemann$^{2}$}

\affiliation{$^1$Department of Physics, Princeton University, Princeton, NJ 08544}

\affiliation{$^2$Max-Planck Institute for the Physics of Complex Systems, D-01187 Dresden, Germany}

\date{\today}

\begin{abstract}
The nature and the very existence of the resonant plaquette valence bond state that separates the classical columnar phase and the Rokhsar-Kivelson point in the quantum dimer model remains unsettled. Here we take a different line of attack on this model, and on the closely related six-vertex model, by exploiting the global conservation law of the number of electric field lines. This allows us to study a single fluctuating electric field line which we show maps exactly onto a one-dimensional spin chain. In the case of the six-vertex model, the electric field line maps onto the celebrated spin 1/2 XXZ model which can be solved exactly. In the quantum dimer model, the electric field line is mapped onto a two-leg spin 1/2 ladder, which we study using
numerical exact diagonalization. Our findings are consistent with the existence of three distinct phases including a Luttinger liquid phase, the one-dimensional precursor to the two-dimensional plaquette valence bond solid. The uncanny resemblance of our quasi-one-dimensional electric field line problem to the full two-dimensional problem suggests that much of the behavior of the latter might be understood by thinking of it as a closely packed array of field lines which themselves are undergoing nontrivial phase transitions.
\end{abstract}
\maketitle

\section{Introduction}

Lattice gauge theories are quantum mechanical lattice models with local conservation laws which often harbor unconventional phases of matter~\cite{fradkin2013field,Wen:2004ym} and can simulate the behavior of the vacuum of our universe~\cite{kogut1979introduction}. Traditionally they have been proposed as descriptions of certain frustrated quantum magnets, such as spin ice materials \cite{Gingras}, but there is also recent interest in alternative platforms such as cold atomic systems~\cite{Zohar,UJW}.

A simple model which realizes a lattice gauge theory is the quantum dimer model (QDM) in the square lattice introduced by Rokhsar and Kivelson (RK) nearly thirty years ago~\cite{RK}. A closely related model is the quantum six vertex model (Q6VM) which can be viewed as a quantum spin ice model in the checkerboard lattice~\cite{Shannon,Moessner2004,rom}. These models fall within the class of quantum link models studied by the lattice gauge theory community~\cite{Chandrasekharan:1996ih,Horn:1981kk}. Both models realize a U(1) lattice gauge theory~\cite{fradkin2013field}, and can be viewed as restrictions to different subspaces of the Gauss law for a common underlying microscopic Hamiltonian, as we will revisit in the next section. This RK Hamiltonian contains a single independent parameter, $v$, that can be tuned to realize different phases. Various features are shared between the QDM and Q6VM. For $v<1$ the ground states in the square lattice are broken symmetry phases, in which the charges are confined from the gauge theory point of view~\cite{Polyakov}. For $v=1$, the Hamiltonian realizes the celebrated RK point, at which a soft quadratically dispersing photon mode appears. For $v>1$ the Hamiltonian has a large number of degenerate states that grows exponentially with the linear dimension of the lattice~\cite{Roderich2011}, and both systems exhibit a form of fragile sub-dimensional deconfinement~\cite{Batista,Shannon}.

\begin{figure}
\label{cartoon}
\centering
 \includegraphics[width=0.9\columnwidth]{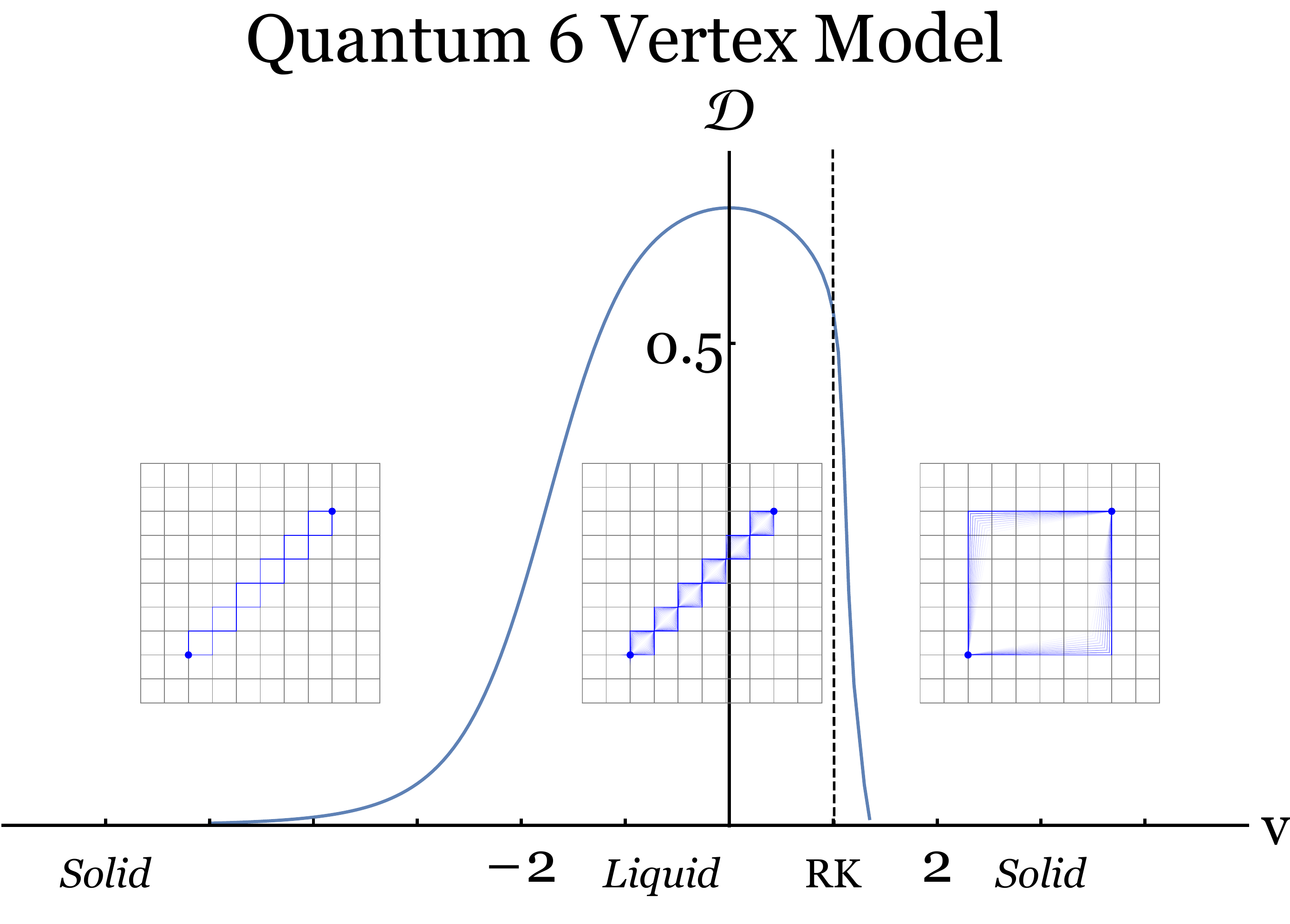}
  \includegraphics[width=0.9\columnwidth]{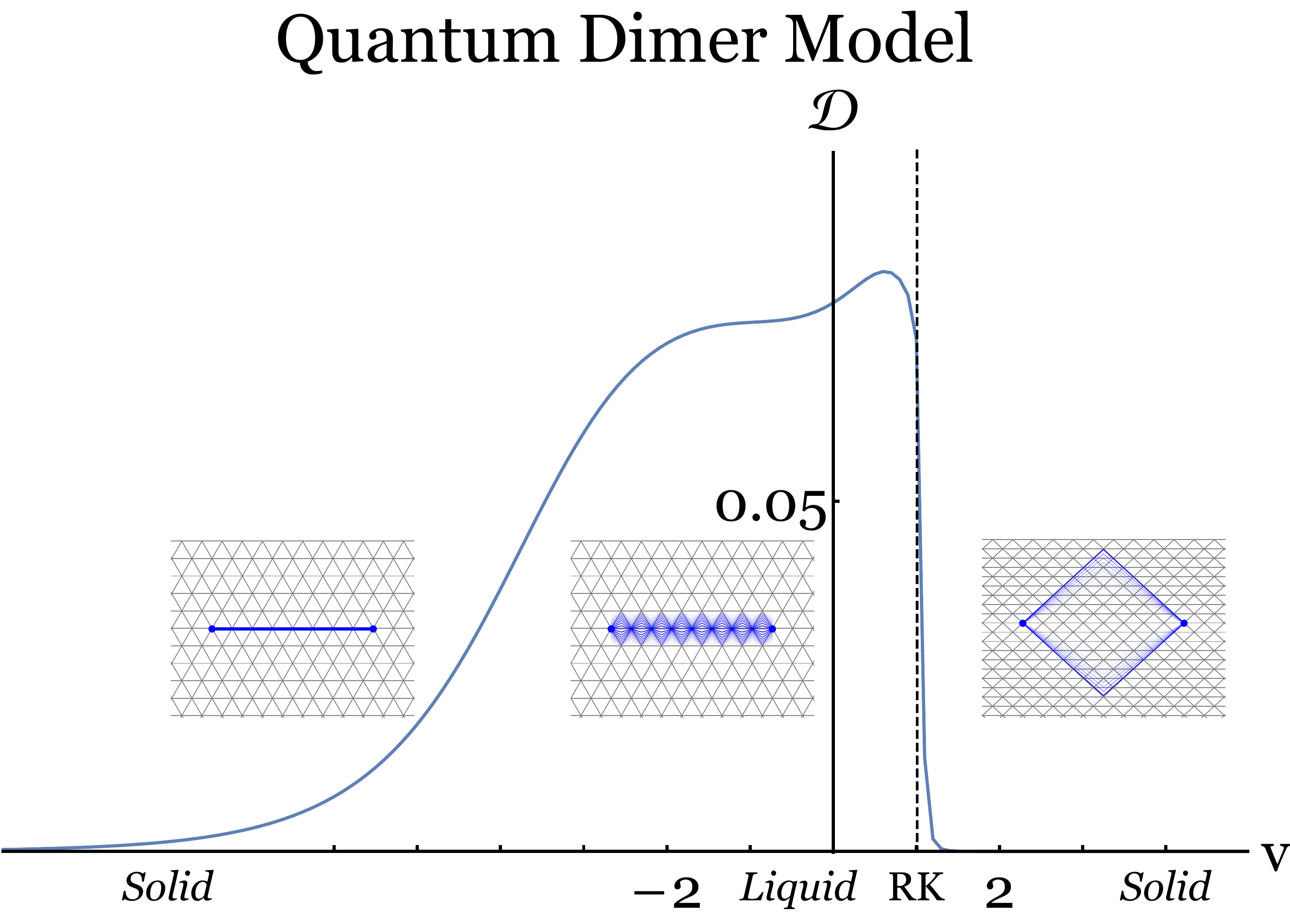}
 \caption{Drude weight and phases for the electric field line for the Q6VM (\textit{top}), which maps onto the spin-1/2 XXZ chain, and the QDM  (\textit{bottom}), which maps onto a spin-1/2 two-leg ladder. Insets depict the ground state of the electric field lines in the different regimes. We see similar trends in Q6VM and QDM, with a fluid state intervening between two solids, although in the latter we rely on exact diagonalization of small systems. The solid phases are precursors to the columnar and staggered phases and the liquid is a precursor to the resonant plaquette phase in two dimensions.}
 \label{cartoon}
\end{figure}
%

Remarkably, in spite of its history, the precise nature of phases of the QDM in the square lattice is still a matter of debate. An early numerical study found evidence for a columnar phase for all $v<1$~\cite{Sachdev}. Two subsequent studies advocated for an intervening plaquette valence bond solid phase separating the columnar phase and the RK point, although they disagreed on its location~\cite{Leung,Plaquette}. A third possibility was put forth in Ref.~\cite{Ralko}, which found evidence for a mixed columnar-plaquette phase intervening between the pure columnar and pure plaquette phases. This study also advocated for a considerably reduced region for the potential pure plaquette phase. More recently, by employing height mappings~\cite{Henley}, two Monte Carlo studies have reached larger system sizes and brought the discussion full circle by advocating that there is no intervening plaquette or mixed phase separating the columnar phase and the RK point~\cite{Banerjee,Garrahan}.

In contrast, the situation is more clear in the Q6VM. An exact diagonalization study~\cite{Shannon} found evidence for a columnar phase for $v \lesssim -0.4$, undergoing a phase transition into a plaquette phase that is stabilized for $-0.4 \lesssim v \leq 1$. Similar conclusions were reached in a subsequent Monte Carlo study employing a height mapping~\cite{Banerjee2013}. This study argued that the transition was first order but it found it to be anomalously weak, which was interpreted as arising from proximity to the terminal tricritical point of the first order phase transition line, although such a putative tricritical point has not yet been explicitly accessed to the best of our knowledge.

In addition to their local conservation laws, which encode the local Gauss law constraint, U(1) lattice gauge theories can have additional global conservation laws that depend on the topology of the space. For example in a 2D torus, there are two topological conserved quantities that measure the total electric flux along the periodic directions in the torus, known as t'Hooft operators~\cite{fradkin2013field}. These two operators, also referred to as winding numbers, can be used to further split the QDM and Q6VM Hilbert spaces into decoupled subspaces or sectors. One can think of these operators as measuring the total number of electric field lines that thread the torus in a given direction. The global ground state of the RK Hamiltonian for $v<1$ belongs to the sector with zero winding in which there are as many electric field lines going to the left, to the right, the up and down directions. For $v>1$, many large winding sectors become degenerate and there is no unique global ground state; a set of exactly degenerate ground states emerges which grows exponentially with the perimeter of the torus, as we will revisit in \Sec{sec:6VM}.\footnote{It is often stated that for $v>1$ the staggered pattern of dimers is the ground state of the RK Hamiltonian, but this is only true if one restricts to a specific winding sector and ignores the degeneracy coming from the sectors.} At the RK point, $v=1$, one indeed has an even larger number of exactly degenerate zero energy ground states. This is because, in addition to the aforementioned ground states that appear in large winding sectors for $v>1$, there appears at least one exact zero energy ground for every winding sector, since at the RK point the Hamiltonian is a sum of projectors. The number of winding sectors in a 2D torus scales as the linear size of the system, thus leading to an additional order $\sim L$ ground states at the RK point.

In the present paper, we study the phase diagram of these models using an unconventional line of attack. We will exploit the conserved winding numbers to isolate the quantum dynamics of a single electric field line. This can be achieved in the torus by restricting to the sectors of the Hilbert space with large winding numbers. Similarly in open boundary conditions, one may study a single electric field connecting two static charges in a background vacuum of inert, fully polarized electric field lines. As we will show, the Hamiltonian governing the quantum mechanics of these isolated electric field lines can be mapped exactly onto one dimensional models. In the case of the six-vertex model, the electric field line maps onto the XXZ spin 1/2 chain. This chain is exactly solvable and its phases are well understood. It has three: a gapped symmetry-breaking anti-ferromagnet, an XY magnet with quasi-long-range order (Luttinger liquid), and a ferromagnet. As we will see, these are, respectively, the natural precursors to the columnar, plaquette, and staggered phases of the two dimensional Q6VM. The RK point corresponds to the critical point between the Ising and XY magnet, and hence has an exact underlying SU(2) symmetry. This will allow us to demonstrate that in the largest winding sectors the RK point has a form of \textit{ perfect charge deconfinement} with exactly zero string tension for finite charge separations. We will also show that the single electric field line of the QDM model maps into a two-leg ladder spin 1/2 chain, which to our knowledge has not been previously studied. We find numerical evidence for three phases in this ladder: a gapped phase, a Luttinger liquid, and a phase-separated state. The Luttinger liquid is the natural precursor of the plaquette phase at zero winding, but we find that it has a substantially reduced Drude weight, compared to the Q6VM case, indicating that the liquid is closer to crystallizing. These results are summarized in \Fig{cartoon} where we show the Drude weight and schematic diagrams of the ground state electric field configurations. We caution, however, that our numerical results are limited to very small system sizes and thus it is possible that richer behavior might arise for the QDM two-leg ladder at larger system sizes.

The resemblance of the quasi-one-dimensional single electric field line problem and the full two-dimensional problem in the zero winding sector suggests that much of the behavior of the latter might be understood by thinking of it as a closely packed array of electric field lines which by themselves are undergoing non-trivial phase transitions. This idea is not unprecedented and was advocated in pioneering work by Orland~\cite{Orland1992,Orland1994}, which, however, incorrectly argued the ground state of the QDM and Q6VM to be a gapless liquid like state. A closely related treatment of the problem of quantum spin ice in the presence of Zeeman fields was developed as well in \Ref{PhysRevLett.108.247210}. A related line of thinking has also been used to study the quantum dynamics of stripes~\cite{Eskes}. In this work, we will follow this line of thinking by focusing on understanding the behavior of a single electric field line in the QDM and Q6VM. In an effort to make our presentation self-contained, we have provided a short introduction to the lattice gauge theory formulations of the QDM and Q6VM in \Sec{sec:structure}. In \Sec{sec:sol} we describe the mappings of the Q6VM and QDM as well as their analytical and numerical characterizations.

\section{U(1) GAUGE STRUCTURE OF THE QDM AND Q6VM}
\label{sec:structure}

In this section we revisit the gauge structure of the QDM and Q6VM. These two models can be viewed as arising from the same underlying microscopic Hamiltonian restricted to two different subspaces of Gauss' theorem. The Hilbert space of this underlying Hamiltonian consists of spin 1/2 degrees of freedom lying on the links of the square lattice. We label lattice sites by $\mathbf{r}=(x,y)$, and assign a rightward orientation to the horizontal links and an upward orientation to the vertical links. The two outgoing links from a given site {\bf r} are labeled by the ordered pair $\mathbf{r},x$ for the rightward link and $\mathbf{r},y$ for the upward link. 

The electric field is viewed as a vector directed along each link, and its orientation relative to that of the link is given by the corresponding  $\sigma_z$ operator:
\begin{equation}
E_{\mathbf{r},x} = \sigma^{z}_{\mathbf{r},x}\;, \qquad E_{\mathbf{r},y} = \sigma^{z}_{\mathbf{r},y}.
\end{equation}
To emulate Gauss' theorem in the lattice, we define a charge operator at every site $Q_{\mathbf{r}}$ as the discrete divergence of the electric field:
\begin{equation}
\nabla \cdot \mathbf{E}_{\mathbf{r}} = 
E_{\mathbf{r},x} - E_{\mathbf{r}-\mathbf{\hat{x}},x} + 
E_{\mathbf{r},y} - E_{\mathbf{r}-\mathbf{\hat{y}},y} \equiv Q_{\mathbf{r}}.
\end{equation}
The charge operators are the generators of the local gauge transformations at every site r, which are explicitly implemented by the action of 
\begin{equation}
\mathcal{U} = \exp \left( i \sum_{\mathbf{r}} \theta_{\mathbf{r}} Q_{\mathbf{r}} \right).
\end{equation}
There are then two essential requirements that any Hamiltonian must satisfy in order to qualify as a gauge theory: it must commute with every $Q_{\mathbf{r}}$ and it should be a sum of local terms.  The gauge transformations acting on the spin raising and lowering operators, $\sigma^{\pm}_{\mathbf{r},\ell}$, as follows:
\begin{equation}
\mathcal{U}\sigma^{\pm}_{\mathbf{r},\ell \,}\mathcal{U}^{\dagger} = 
e^{\left(\pm 2i \mathlarger{\int} \nabla \theta\cdot d\boldsymbol{\ell} \right)}\sigma^{\pm}_{\mathbf{r},\ell} = 
e^{\pm 2i \left( \theta_{\mathbf{r}+\mathbf{\hat{l}}}-\theta_{\mathbf{r}} \right)}\sigma^{\pm}_{\mathbf{r},\ell} .
\end{equation}
Thus, $\sigma^{\pm}_{\mathbf{r},\ell}$ transforms like a charge hopping
(or dipole creation) operator of the form $c^\dag_{\mathbf{r} + \ell} c_{\mathbf{r}}$, endowing these operators with the following notion of directionality:
\begin{eqnarray}
\begin{matrix}

\includegraphics[scale=1]{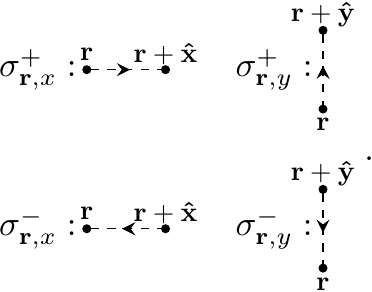}

\end{matrix} .
\end{eqnarray}
More generally we can introduce a charge hopping operator associated with a directed line $\gamma$ that starts at site ${\bf r}$ and ends at site ${\bf r'}$, by taking ordered products of $\sigma^{\pm}_{\mathbf{r},\ell}$ following the natural convention for a discrete line integral, as follows:
\bea
\label{eq:chargehopping}
L_{\pm} &= \prod_{\mathbf{r} \in \gamma} \sigma^{\pm}_{\mathbf{r}} \ .
\eea
The above charge hopping operator for open lines is not gauge invariant. Gauge invariant operators can be obtained by taking ordered products of $\sigma^{\pm}_{\mathbf{r},x(y)}$ to form oriented closed loops, which guarantees that the phases from gauge transformations cancel in pairs. The smallest non-trivial closed loop, associated with each plaquette, is:
\begin{equation}
\label{eq:Magnetic-loop-contract}
\begin{matrix}
\includegraphics[scale=1]{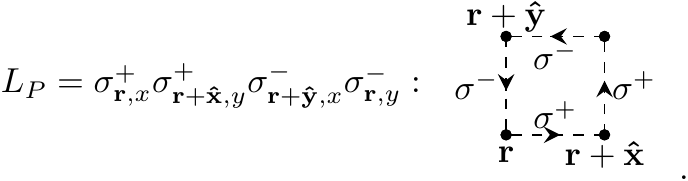}
\end{matrix}
\end{equation}
We will sometimes refer to these closed loop charge hopping operators as magnetic loop operators.

As we have described, the divergence of the electric field is a locally conserved quantity in a U(1) lattice gauge theory. This conservation law expresses a dynamical conservation of the number of oriented electric field lines pierce any given region bounded by a contractible loop. This can be made more explicit by making use of the lattice version of Gauss's theorem, which reads as follows:
\begin{equation}
\int_{S} \nabla\cdot \mathbf{E} = \sum_{\mathbf{r}\in S}\nabla\cdot \mathbf{E}_{\mathbf{r}}
 = \sum_{\mathbf{r}\in \partial S} \mathbf{\hat{n}}\cdot \mathbf{E}_{\mathbf{r}} = \oint_{\partial S} (\mathbf{\hat{n}}d\ell) \cdot \mathbf{E}_{\mathbf{r}}
\end{equation}

\begin{figure}
\centering	
\includegraphics[scale=1]{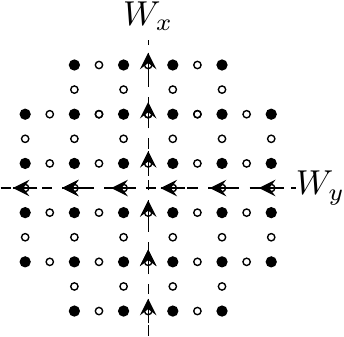}
\caption{The links intersected by the above two non-contractible loops on the torus are those that are summed over to form the 't Hooft operators described in Eq~\eqref{eq:thooft}. }
\label{fig:tHooft-operators}
\end{figure}

It is convenient to transform the integral above into a conventional line integral. For this purpose we define a \textit{dual electric field} $\mathbfcal{E} = \textbf{E} \times \hat{z}$, which is a rotated version of the electric field $\mathbf{E}$. Since it is rotated, it is often convenient to visualize it as residing on the links of the dual lattice. In this way, Gauss' theorem law for a region bounded by a contractible loop can be expressed as a conventional line integral of the dual electric field
\bea
\int_{S} \nabla\cdot \mathbf{E} &=  \oint_{\partial S} (d \ensuremath{\boldsymbol\ell} \times \hat{\mathbf{z}}) \cdot \mathbf{E}_{\mathbf{r}} = \oint_{\partial S} d \ensuremath{\boldsymbol\ell} \cdot \mathbfcal{E}
\eea
Therefore, any closed line integral of the dual electric field over a contractible oriented loop in the dual lattice is a constant of motion. Now, in geometries with non-trivial topology such as the torus, it is possible to have non-contractible loops for which Gauss' theorem as described above cannot strictly be applied. However, it is easy to show that provided that the underlying Hamiltonian is local and also gauge invariant (i.e. that it commutes with the divergence of the electric field at every site), then the line integral of the dual electric field over such noncontractible loops necessarily commutes with the Hamiltonian as well \cite{fradkin2013field}. These operators are known as t'Hooft operators or winding operators. In a torus there are two independent t'Hooft operators associated with the two principal non-contractible loops, defined as:

\begin{equation}
\label{eq:thooft}
\begin{split}
W_{x} &= \oint +d\ell_{y}E_{\mathbf{r},x} = \sum_{\uparrow}E_{\mathbf{r},x}
 = \sum_{\uparrow}\sigma^{z}_{\mathbf{r},x}	\\
W_{y} &= \oint -d\ell_{x}E_{\mathbf{r},y} = \sum_{\leftarrow}E_{\mathbf{r},y}
 = \sum_{\leftarrow}\sigma^{z}_{\mathbf{r},y}
\end{split}
\end{equation}
where $x,y$ designates the direction of the electric lines which are integrated (see \Fig{fig:tHooft-operators}). We will refer to the eigenspaces with definite t'Hooft operators as winding sectors. Therefore, a U(1) lattice gauge theory has additional topological conservation laws which are distinct from the local conservation laws associated with charge conservation. These global conservation laws are associated with a gauge group $\mathrm{U(1)}_{x}\otimes \mathrm{U(1)}_{y}$ which acts as:

\begin{equation}
\mathcal{G}(\theta_{x},\theta_{y}) = \exp\left(
i \theta_{x}W_{x} + i \theta_{y}W_{y}
\right) .
\end{equation}

Notice that charge hopping line operators as defined in \Eq{eq:chargehopping} commute with t'Hooft operators only if they intersect an even number of times because $[\sigma^z, \sigma^{\pm}] = \mp \sigma^\pm$, so each adds a term to the commutator of alternating sign. 

This implies that any closed contractible magnetic loop, such as that from \Eq{eq:Magnetic-loop-contract} , commutes with the t'Hooft operators. However, this is not the case for closed non-contractible magnetic loop operators (Wilson loop operators) and for open line charge hopping operators which intersect the t'Hooft operator an odd number of times. 

The commutator between the t'Hooft operators and one such charge hopping or Wilson loop operator, $L$, is:
\bea
\null [W_{x'}, L ] &= [ \sigma^z_{\mathbf{r}',x'}, \sigma^\pm_{\mathbf{r}',x'}] \prod_{x \neq x'} \sigma^\pm _{\mathbf{r},x} = \pm L ,
\eea

where the $\pm$ sign is dictated by the direction of the line integral used to obtain $L$. Therefore, the Wilson loop operators act as raising and lowering operators for the t'Hooft operators, and move between different winding sectors of the Hilbert space. The action of the Wilson loop operators can be interpreted as the process of creating a pair of opposite charges from the vacuum and annihilating them after hopping them over a non-contractible loop, which results in change in the winding number.

\subsection{Construction of the Hamiltonian}
The structure described above is generic to $\mathrm{U}(1)$ lattice gauge theories in 2+1D. In this section we will describe a specific Hamiltonian that reduces to the RK Hamiltonians of the QDM and Q6VM in the appropriate subspaces of the Gauss law. The Hamiltonian reads
\begin{equation}
\label{eq:ham}
H = -t \sum_{P} \left( L^{\textcolor{white}{\dagger}}_{P}+L^{\textcolor{black}{\dagger}}_{P} \right)
 + V \sum_{P} \left( L^{\textcolor{black}{\dagger}}_{P}L^{\textcolor{white}{\dagger}}_{P}
 + L^{\textcolor{white}{\dagger}}_{P}L^{\textcolor{black}{\dagger}}_{P} \right) \ .
\end{equation}
where the plaquette operator $L_{P}$ is defined in Eq.~\eqref{eq:Magnetic-loop-contract}. The first term is the analogue of the magnetic field term in Maxwell theory which induces quantum fluctuations of the electric field configurations. The second term is diagonal in the electric field configurations and acts as a potential. In fact,
\begin{equation}
L^{\dagger}_{P}L_{P} = \resizebox{.8\hsize}{!}{$\left(\dfrac{1-\sigma^{z}_{\mathbf{r},x}}{2}\right) \left(\dfrac{1-\sigma^{z}_{\mathbf{r}+\mathbf{\hat{x}},y}}{2}\right) \left(\dfrac{1+\sigma^{z}_{\mathbf{r}+\mathbf{\hat{y}},x}}{2}\right) \left(\dfrac{1+\sigma^{z}_{\mathbf{r},y}}{2}\right)$}
\end{equation}
Thus we see that $L_{P}^{\dagger}L_{P}$ projects onto a loop eigenstates of the electric field at a given plaquette. These two configurations are depicted in Fig~\ref{fig:Six-vertex-Path-1} and are denoted by $\ket{\circlearrowleft}$ and $\ket{\circlearrowright}$.  If a plaquette has either of these two electric field configurations, we call it flippable (otherwise, not flippable). Notice that the Hamiltonian in \Eq{eq:ham} only acts on flippable plaquettes; non-flippable plaquettes are annihilated by both the magnetic and the electric field terms. Therefore, the Hamiltonian can be expressed in the following more intuitive notation:
\bea
\label{eq:ham2}
H &= \sum_{\Box} -t (\ket{\circlearrowleft} \bra{\circlearrowright} + \ket{\circlearrowright} \bra{\circlearrowleft}) + V (\ket{\circlearrowleft} \bra{\circlearrowleft} + \ket{\circlearrowright} \bra{\circlearrowright} ).
\eea
Notice that the term proportional to $V$ counts the total number of flippable plaquettes and the $t$ term exchanges flippable plaquettes. There is a single parameter $v \equiv V/t$ which controls the physics. The  RK point is located at $v= 1$ where the Hamiltonian becomes a sum of projectors. 

\subsection{Six-vertex model}

\begin{figure}
\centering	
\includegraphics[scale=1]{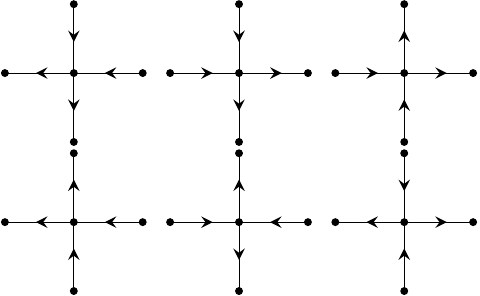}
\caption{Six vertices consistent with $Q_{\mathbf{r}}=0$.}
\label{fig:Six-vertex-model}
\end{figure}

The six-vertex model is realized in the subspace in which the charge operator is set to zero, $Q_{\mathbf{r}} = 0$, for every site $\mathbf{r}$, which constrains the electric fields at each site to one of the six configurations shown in Fig. \ref{fig:Six-vertex-model}, and hence the name of the model. Additionally, the Hilbert space further separates into winding sectors specified by the t'Hooft operators $W_{x,y}$ from \Eq{eq:thooft}. Let us consider the sector with maximal winding: $W_x = L_x, W_y = L_y $ where $L_{x,y}$ are the number of sites along the $x,y$ directions. This maximal winding sector contains a single eigenstate in which the electric field at every link is $\sigma_{\mathbf{r}, x(y)} = +1$. Such state is trivially a zero energy eigenstate of the Hamiltonian in \Eq{eq:ham2} as it contains no flippable plaquettes and is one of the exponentially large number of degenerate ground states realized to the right of the RK point for $v>1$. We will see this trivial sector as the reference vacuum on top of which we will create a single electric field line by the action of a charge hopping operator $L$ from \Eq{eq:chargehopping} along a suitably chosen path.

\begin{figure}
\centering	
\includegraphics[scale=1]{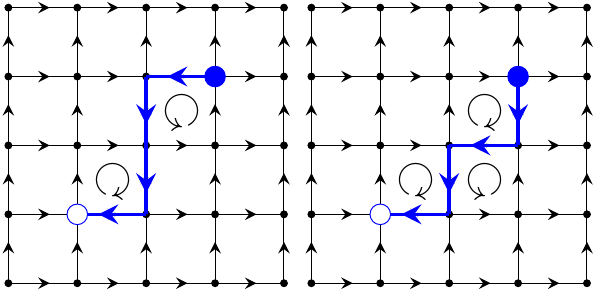}
\caption{One possible path on the lattice before (left) and after (right) having acted with $L_{P}^{\dagger}$ on the counterclockwise flippable plaquette of the figure on the left.}
\label{fig:Six-vertex-Path-1}
\end{figure}

Specifically, we act on the reference vacuum with a charge hopping operator $L$ which reverses the electric field along this path and terminates at two sites where charges $Q_\mathbf{r}=\pm 2$ are created. This state then evolves under the Q6VM Hamiltonian, \Eq{eq:ham2}. An example of the action of $H$ is depicted in Fig. \ref{fig:Six-vertex-Path-1}, in which it is shown how a plaquette flips with a corresponding change in the electric field line configuration. As we will show in \Sec{sec:xxzmap}, this problem maps identically into the spin 1/2 XXZ chain.
\label{sec:6VM}

\subsection{Quantum dimer model}
The QDM model is realized in a different subspace of the Gauss law. This is achieved by splitting the sites of the square lattice into two sub-lattices and requiring:
\bea
Q_{\mathbf{r}} &= 2(-1)^{r_x + r_y}.
\eea
We call the positive (negative) charge sub-lattice A (B).  There are four allowed configurations for the electric field lines in the links touching a site in either the A or B sublattices. For the A sublattice, three electric field lines always flow into the site, while a single field line goes out, and the reverse holds for the sublattice B. We can denote the configurations by covering the unique link where the electric field flows into the A sublattice with a dimer. This is also the unique link which flows out of the B sublattice. Since every lattice site is touched by one and only one of these dimers, the allowed configurations of electric field lines in this subspace are in one to one correspondence to arrays of closely packed non-overlapping dimers that fully cover the lattice. 

Similarly to the Q6VM, the reference vacuum is chosen so that there are no flippable plaquettes and is a state with an staggered array of dimers such as that depicted in \Fig{fig:dimer}. Notice that this state has winding numbers for the t'Hooft operators given by $W_x = 0,W_y = L_y$. Therefore, unlike the Q6VM, within the subspace of the QDM model, it is impossible to simultaneously maximize the winding numbers along both directions of the torus.

This staggered vacuum is an exact zero energy eigenstate of the RK Hamiltonian from \Eq{eq:ham2}. We can again study how the addition of charges alters the reference vacuum. As we will show in \Sec{sec:mapdimer}, we can act on the background with a open charge hoping line operator. If the path satisfies certain simple conditions, charges will only be created at the ends of the line. The resulting state will now be dynamical, and, as we will see, it maps onto a two-leg ladder of spin 1/2 degrees of freedom. In \Fig{fig:dimer}, we show an example of such a state and the action of the Hamiltonian. 

\section{Solution of the Models}
\label{sec:sol}

In this section, we consider the problem of a pair of charges connected by a single fluctuating electric field line. We will map these problems to the XXZ chain for the Q6VM and a two-leg ladder for the QDM. The phase diagram is known in the former case, and we will provide numerical evidence that the diagram is the qualitatively similar for the latter. 

\subsection{Mapping the electric field line of the Q6VM to the XXZ chain}
\label{sec:xxzmap}

\begin{figure}
\centering	
\includegraphics[scale=1]{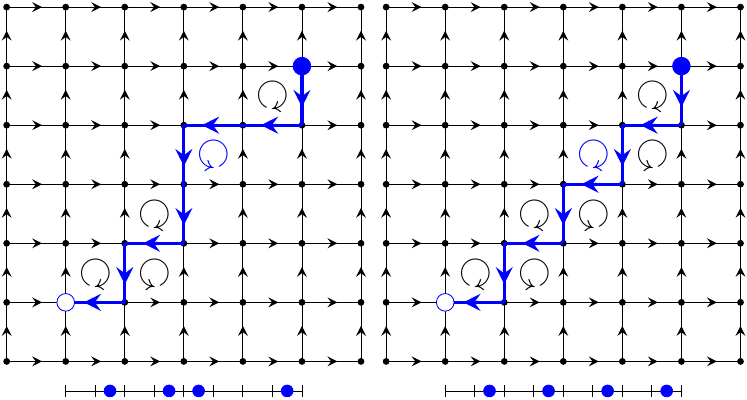}
\caption{\textit{Top:} Two states of a string on a square lattice. \textit{Left: } The state at $t=0$ has five flippable plaquettes and is an eigenstate with energy $5V$. \textit{Right:} The state has seven flippable plaquettes, and is an eigenstate too, with energy $5V$. This is also a maximally kinked state.
\textit{Bottom:} The equivalent configuration of the hardcore boson chain. }
\label{fig:path}
\end{figure}

We construct a state with a single electric field line on top of the fully polarized state, which we call the reference vacuum. This reference vacuum is the unique state where the electric field is positive along every link of the lattice. We then add two charges $\pm2$ to prepare the electric field line. Preservation of the local Gauss law requires that the two charges must be connected by a sequence of contiguous links on which we reverse the electric field relative to the reference vacuum. Notice that a flippable plaquette appears at every corner of the electric field line, and therefore such corners have non-trivial quantum dynamics under the action of the Hamiltonian, \Eq{eq:ham2} as depicted in \Fig{fig:path}. Since the reversed electric field line must flow against the polarization of the reference vacuum, it is bounded by the rectangle whose corners are defined by the charges. We take one charge to be sitting at the origin and the other at the site ${\bf r}=(\ell_x, \ell_y)$. The constraint that the line must run against the polarization of the vacuum gives rise to a dynamical conservation of its length, measured using the taxicab distance $|\ell_x| + |\ell_y|$. An analogous dynamical constraint appears in the strong coupling expansion of lattice QCD \cite{RevModPhys.55.775,PhysRevD.23.2945}. 

To map this fluctuating electric field line onto a conventional $1D$ system, we unfold the line into a chain of $L\equiv \ell_x + \ell_y$ sites labeled by $i$. We fill the chain with hardcore bosons $b^\dag_i$ defined by
\bea
\null [b_i, b_j] &= 0, \quad  i \neq j  \\
\{ b_i, b^\dag_i \}&= 1 \ ,\\
\eea
so that every horizontal segment corresponds to an empty site and every vertical segment corresponds to a filled site. Because both $\ell_x$ and $\ell_y$ are conserved quantities, the system has a global $U(1)$ symmetry generated by the conservation of the total particle number:
\bea
N_b &= \sum_i b^\dag_i b_i = \ell_y .
\eea
This allows us to define a filling $\nu = \frac{N_b}{L} = \frac{1}{1+\ell_x/\ell_y}$.

To write the Hamiltonian of the boson chain, we note that the action of the kinetic term in \Eq{eq:ham2} is to flip a corner, reversing the order of its constituent horizontal and vertical segments. This corresponds to hopping a boson between nearest neighbor sites. The potential term counts the number of corners. Therefore these corners can be viewed as counting the links of the 1D lattice on which there is a change of the absolute value of the occupation number of the bosons. Thus, the Hamiltonian can be written as:
\bea
\label{hline}
H_{6v} &= \sum_{i=1}^{L} -t (b^\dag_i b_{i+1} + h.c.) + V (n_i - n_{i+1})^2 , \\
\eea
Note that we may fix $t>0$ because its sign may be changed via a unitary transformation $b^i \to (-1)^i b_i$. The single electric field line has therefore a Hamiltonian which is equivalent to a 1D Bose-Hubbard model with nearest neighbor hopping and interactions. This model has a particle-hole symmetry $\Theta$:
\bea
\Theta^\dag b_i \Theta &= b^\dag_i  \\
\Theta^\dag b^\dag_i \Theta &= b_i  \\
\eea
which takes $\Theta^\dag n_i \Theta = 1 - n_i$. The ground state is only invariant under this symmetry for $\nu = \frac{1}{2}$. This particle-hole conjugation can be interpreted as the a reflection of the electric field line configuration along the diagonal of the square lattice that intersects the charge located at the endpoint of the electric field line at the origin $(0,0)$ and therefore swaps the coordinates of charge at the end-point located at $(l_x,l_y)$, $l_x \leftrightarrow l_y$. Therefore the ground state is invariant only for charges displaced diagonally on the lattice. 

\begin{figure}
\begin{center}
\includegraphics[width=8cm]{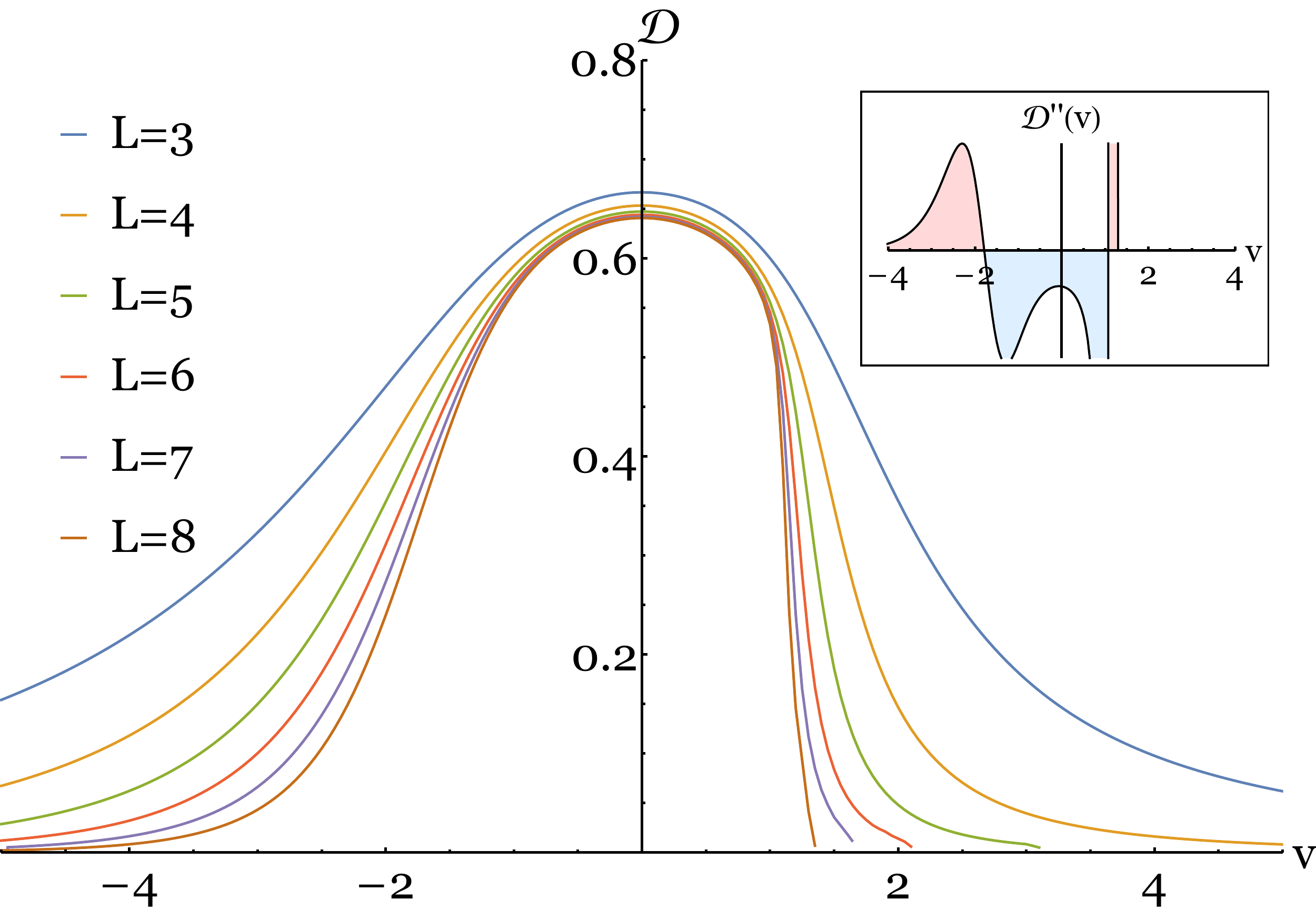} 
\includegraphics[width=8cm,trim=1.5cm 0 0 0]{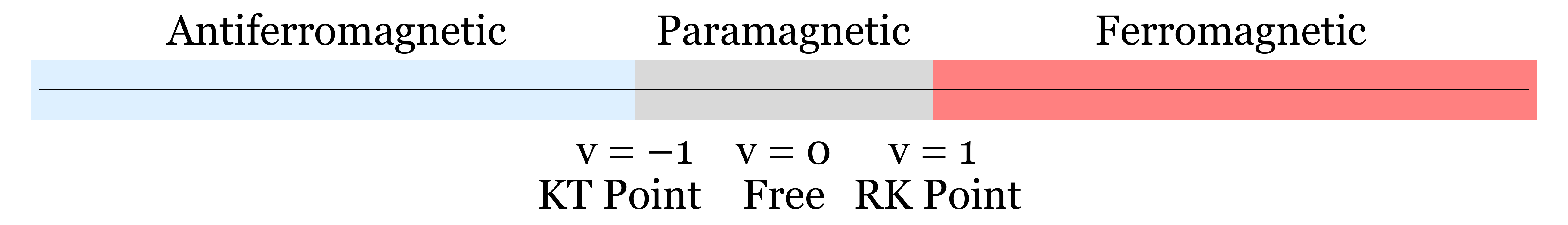} 
\caption{\textit{Top}: The Drude weight of the model is computed for systems of size $L= 3,\dots, 8$ using exact diagonalization. The phase transition at the RK is obvious as $\mathcal{D} \to 0$ at $v = 1$, but the convergence is slower for the infinite order Kosterlitz-Thouless transition at $v = -1$. The inset shows the second derivative of the Drude weight $\frac{d^2 \mathcal{D}}{d^2 v}$ which changes sign roughly at the phase transitions. \textit{Bottom}: The exact phase diagram is shown at $M=0$.}
\label{fig:XXZ}
\end{center}
\end{figure}

As is well known, this model can be mapped from hard-core bosons onto spin 1/2 degrees of freedom. We let $b_i = S_i^-,  b^\dag_i = S_i^+, n_i = S^z_i + \frac{1}{2}$ and find
\bea
H_{6v} &= -J \sum_{i=1}^L \lp S_i^x S_{i+1}^x + S^y_i S^y_{i+1} + v S^z_i S^z_{i+1} - \frac{v}{4} \rp
\eea
where $J = 2t, v = V/t$. This Hamiltonian generalizes the one first obtained by Orland in \Ref{Orland1992} to non-zero values of $v$. This is the well-known XXZ spin chain which is an exactly solvable model \cite{Yang:1966ty, PhysRev.150.327}. At zero magnetization, or half filling, there are three phases for XXZ model: a gapped Ising ferromagnet for $v>1$, a gapless XY-like magnet with power law correlations for $|v|<1$, and a gapped Ising anti-ferromagnet for $v<-1$, as illustrated in \Fig{fig:XXZ}. The energy density of these phases, defined as the total energy divided by the total number sites, plays the role of the tension of the electric field line and dictates the interaction law between the charges. In general it can be written as:
\bea
E(v,M) &= f(v,M) L + O(L^{-1}), \\
M &= \frac{2}{L} \sum_i S^z_i = \frac{\ell_y - \ell_x}{L}
\eea
where $M$ is the conserved magnetization and $f(v,M)$ is the free energy density. \Ref{PhysRev.150.327} contains expressions for $f(v,M)$ obtained from the Bethe Ansatz. We will included selected results from this work in the following more detailed description of the phases, and the reader is otherwise referred to the original paper.

We begin in the $v > 1$ ferromagnetic phase at fixed $M$. The global ground state is the trivial fully polarized state with $|M|=1$ and energy $E=0$. For $|M| \neq 1$, the system is unable to reach the global fully-aligned ground state due to the conservation of $M$.  Instead, a domain wall is created, corresponding in the bosonic picture to phase separation and in the electric field line picture to the path approaching its bounding rectangle. When this domain wall is large compared to the lattice scale (namely when $v \to 1$), it can can be modeled using a semi-classical continuum description. Using the $\mathrm{U}(1)$ symmetry, we can fix the spin to lie in a plane with $S^z_i \to \sin \theta(x),S^x_i \to \cos \theta(x)$. In this approximation, the energy is given by
\bea
E &= \int_0^L dx \lp \frac{\rho}{2} \lp \frac{d \theta}{dx} \rp^2 - \la \cos^2 \theta \rp \\
\eea
with $\rho = J , \la = J(v - 1)$ in units where the lattice spacing is one. One can minimize this energy using Euler-Lagrange equations, and they may be exactly solved in terms of Jacobi amplitudes for finite $L$. In the limit $L \to \infty$ the solution is given by
\bea
S^z(x) &= \tanh \frac{x-x_0}{w}, \quad w = \frac{1}{\sqrt{2(v -1)}}
\eea
where $w$ is the width of the domain wall and $\frac{x_0}{L} = \frac{1-M}{2}$ fixes the magnetization. This solution has $\theta \sim 0, \pi$ except in the region $|x-x_0| \lesssim w$, so the presence of a domain wall contributes an energy $O(L^0)$ to the system, which can be estimated to leading order in $v-1$ to be:
\bea
E &=2 \sqrt{2} J \sqrt{v -1} \\
\eea

This result is noteworthy as it encodes a form of subdimensional charge deconfinement emerging in the staggered phase to the right of the RK point for $v>1$~\cite{Shannon,Batista}. Namely, when two charges are separated strictly along the horizontal and vertical directions of the square lattice they have an electric field line that costs zero energy and are therefore deconfined; however, when the charges try to move perpendicularly, a domain wall is created that incurs in an energy cost $2 \sqrt{2} J \sqrt{v -1}$. This form of subdimensional deconfinement differs however from the recently studied case of fracton models (see e.g \cite{Chamon,Haah,Yoshida,Vijay,Pretko}).

A remarkable implication of our mapping is that an $\mathrm{SU}(2)$ symmetry emerges at the RK point, since it occurs at $v = 1$ where the XXZ chain becomes the isotropic ferromagnetic Heisenberg model. This has profound consequences on charge confinement. Since the exact ground state of the ferromagnetic is a simple tensor product of spins pointing in the same direction, the exact ground state energy is $E=0$. Because the ferromagnet can smoothly cant away from the $z$ axis without any energy cost due to the exact $\mathrm{SU}(2)$ symmetry, the electric field has vanishing tension independent of orientation of the charges. One should bear in mind that the background we are considering has exactly zero energy at the RK point, so it belongs to the full zero energy manifold even when the restriction of global winding numbers is removed. Recall that at the RK point, all states have non-negative energy because the Hamiltonian is a sum of projectors. Therefore, we have proven that there is ground state at the RK point with exactly zero string tension even for finite charge separations. This is a form of \textit{perfect} charge deconfinement, which contrasts with the generic deconfinement in which the vanishing string tension appears in the asymptotic thermodynamic limit of large charge separations compared to the lattice scale.

The system then enters a gapless phase with power law correlations for $|v| < 1$, undergoing a continuous spin-flop type transition from an Ising magnet with easy axis anisotropy to an XY magnet with easy plane anisotropy through an SU$(2)$ invariant point. The kinetic term dominates and the plaquettes are strongly resonating. Much of the behavior of this phase can be understood by studying the point $v = 0$, at which the Hamiltonian may be mapped into free $1D$ fermions with the Jordan-Wigner transformation. This is most easily seen from \Eq{hline} where the interaction term vanishes. It is then simple to obtain
\bea
f(0, M) &= -\frac{J}{\pi} \cos \frac{\pi}{2} M \\
\eea
This function is mimized at $M=0$ where the electric field line is oriented diagonally. We will see later that this is also the preferred orientation of the antiferromagnet.


Because of its strong plaquette flipping fluctuations (which correspond to boson hopping processes), the present gapless phase can be seen as a precursor to the resonating plaquette phase of the full 2D zero winding sector of the Q6VM. Such zero winding sector can be seen as containing a finite density of electric field lines which we have studied. The electric field lines have, however, strong interactions, and these are presumably responsible for turning the liquid Luttinger phase that we encountered into the plaquette crystalline phase seen in numerical studies~\cite{Shannon,Banerjee}.

\begin{figure}
\begin{center}
\includegraphics[width=8cm]{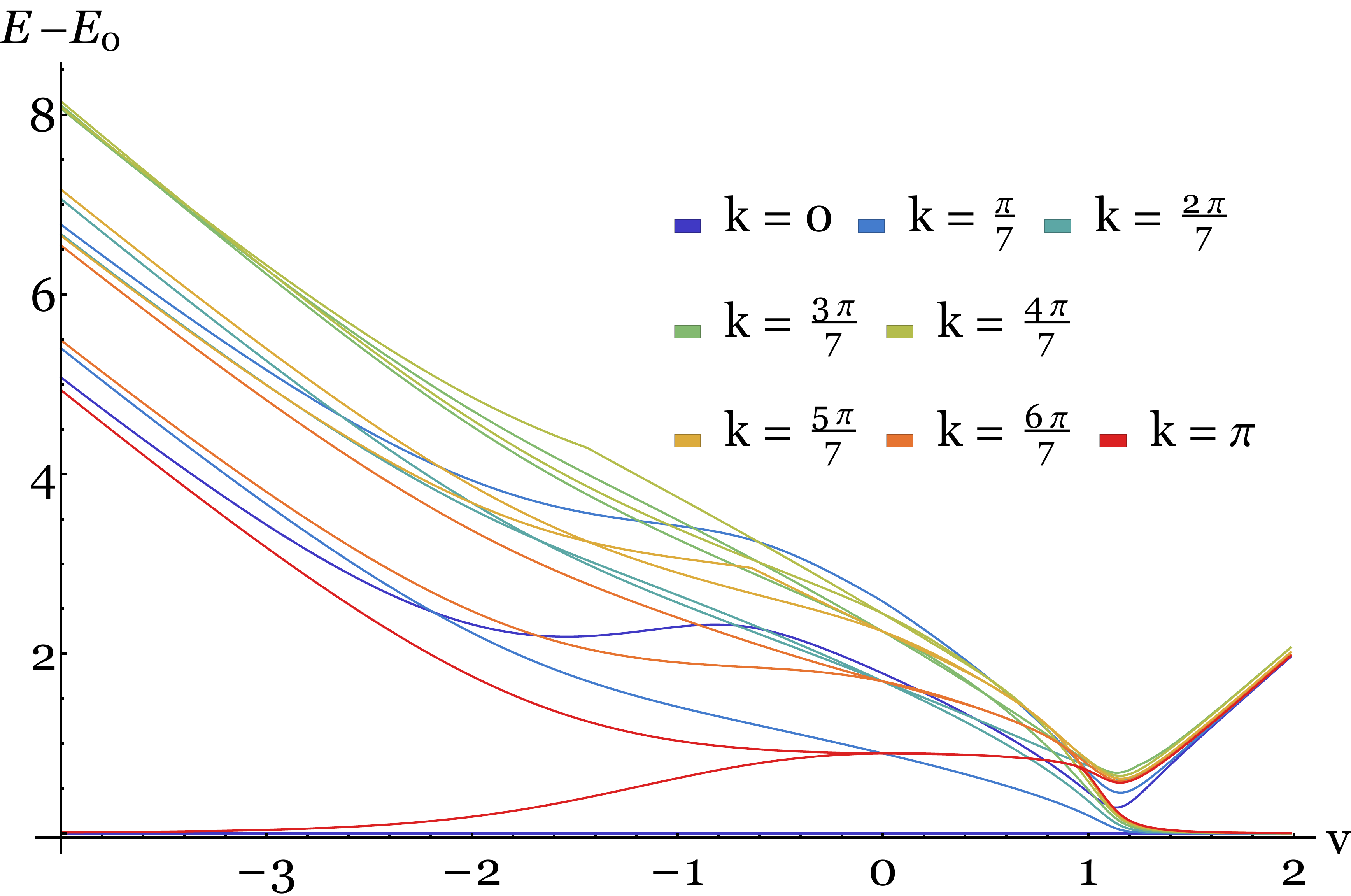} 
\caption{The two lowest energy states in each $k$ sector for $L=7$ are shown relative to the ground state energy for $\phi =0$. At large negative $v$, the odd and even charge density waves at $k=0, \pi$ become degenerate, and the low lying excitations are propagating domain walls. The discontinuity in the first derivative at the RK is visible. On the right, the lowest energy state at each $k$ become approximately degenerate.}
\label{fig:Q6Vspec0}
\end{center}
\end{figure}

At $v =-1$, the system undergoes a form of the Kosterlitz-Thouless transition and the free energy is smooth to all orders \cite{PhysRev.150.327} \footnote{An SU$(2)$ symmetry also exists at $v=-1$, as can be seen after a rotating the spins on alternating sites to flip the sign of the $XY$ terms. Thus, $H_{v=1} = - H_{v=-1}$. As a result, the ground state at $v=-1$, a spin singlet, is the highest excited state at $v=1$, where the ground state is fully polarized. } . In \Fig{fig:Q6Vspec0}, we compute the lowest two energy levels at each momentum for a finite sized system; the RK point is clear in the spectrum at $v=1$ but the transition at $v=-1$ is much less so. The gap reopens for $v < -1$ and the system becomes antiferromagnetic for the rest of the parameter space, corresponding in the boson language to a charge density wave. In this regime, the path approaches the diagonal, becoming jagged, and quantum fluctuations freeze out as $|v|$ increases. 

It is instructive to compare the behavior of the model with explicit numerical solutions. Although this is not strictly needed in the present case where the phases can be understood exactly, it will prove essential for the QDM model where we lack exact solutions. To diagnose and distinguish fluid and gapped phases, it is useful to introduce the Drude weight $\mathcal{D}$. The Drude weight is finite in the gapless Luttinger liquid phase in which the bosons can flow in response to a probe electric field and vanishes in the insulating phases. The Drude weight can be conveniently computed by twisting the boundary conditions of a length $L$ one-dimensional chain~\cite{Kohn}, which we implement by adding a phase $e^{i\phi}$ at the last bond as  follows: \footnote{Notice that this is still a translationally invariant problem with a conserved momentum as the flux can be spread uniformly by changing the gauge via a unitary transformation $b^\dag_j \to e^{ i\frac{j\phi}{L}} b^\dag_j$, which leads to a complex hopping amplitude $t e^{i \phi/L}$. } 

\bea
H_{6v, closed} &= H_{6v, open} + e^{i \phi} t b^\dag_1 b_L + h.c. \ . 
\eea

\noindent The eigenvalues of the Hamiltonian can be determined as functions of $\phi$ numerically and the Drude weight can be computed as the stiffness for twisting boundary conditions~\cite{Kohn}:
\bea
\mathcal{D} \equiv L \frac{d^2 E_{0}}{d\phi^2} \ .
\eea
In \Fig{fig:XXZ}, we see that the Drude weight distinguishes the conducting and insulating phases fairly well even for modest $L$. We will find a similar behavior for the QDM. 

\subsection{Mapping the Electric Field line of the QDM to a two-leg ladder}
\label{sec:mapdimer}

\begin{figure}
\centering	
\includegraphics[scale=1]{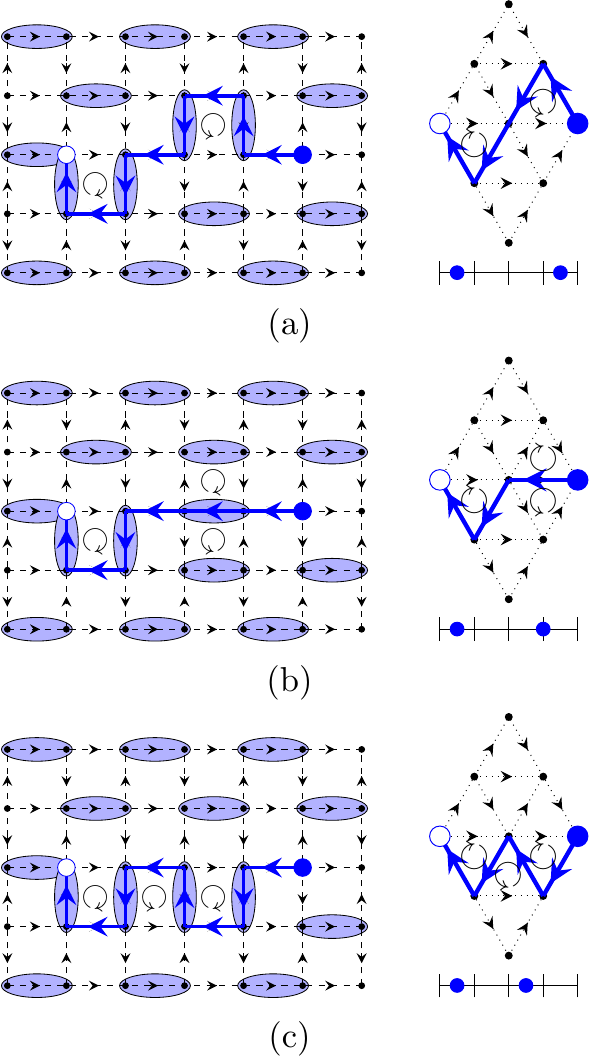}
\caption{\textit{Left:} Figures (a), (b) and (c) show succesive states of an open string after having acted with the Hamiltonian on the right flippable plaquette. \textit{Right:} The same configurations and shown in the triangular lattice where the dimers are contracted to points.\textit{Bottom:} The projection into a $1D$ chain is shown, including the flipping from diagonal to horizontal configurations as hoppings from odd to even sites. }
\label{fig:dimer}
\end{figure}

Having studied the Q6VM, we now search for a similar mapping in the QDM. To create a single electric field line, we follow an analogous procedure. Without loss of generality, we choose the line to start and end on the A sublattice sites with charges $Q_{\mathbf{r}} = +2$~\footnote{When the path terminates in the B sublattice the last link remains inert under the action of the Hamiltoninan.}. We draw a directed line through a sequence of connected links always against the flow of the background configuration. Notice that flowing against the reference vacuum requires the path to follow a sequence of links which alternates between those with dimers and those without. The line is constructed so that defect monomer charges appear only at the ends of the line, with values $Q_{\mathbf{r}0} = 0$ at the starting site and $Q_{\mathbf{r}f} = +4$ at the ending site, as illustrated in \Fig{fig:dimer}. Along this path, we reverse the direction of the electric field compared to the reference vacuum. This causes the dimers initially present in the background configuration to become unoccupied, and  the initially unoccupied links of the path to now contain dimers. At the ends of the line, two dimers overlap the first site and no dimers touch the last.

To map the problem into a conventional 1D system, it is convenient to simplify the picture by representing the dimers of the background configuration as points. In this way the electric field line appears as a directed string joining the sites of a triangular lattice as depicted in \Fig{fig:dimer}. In this picture, a single flippable (triangular) plaquette is created when segments of opposite slope are joined into a triangular corner, and \textit{two} are created when the segment is horizontal. No flippable plaquettes are created when a horizontal segment joins a sloped segment. This dictates the form of the diagonal potential term. The kinetic term locally changes the electric field line by flipping triangular corners to horizontal segments. One such possibility is shown in \Fig{fig:dimer} (b). As demonstrated there, string's length along the triangular lattice is not conserved, making mapping to a 1D chain less straightforward. Notice that in this case the string is also constrained to move within a finite region determined by the position of the charges similarly to the Q6VM. In the triangular lattice picture this region is a parallelogram as illustrated in \Fig{fig:dimer}.

We have found, however, a mapping onto 1D hard-core bosons moving in a chain with a basis of two distinct sites which can be viewed as an asymmetric two leg ladder. This chain is obtained by projecting the triangular lattice on a horizontal line to make a one-dimensional lattice as depicted in \Fig{fig:dimer}. This one-dimensional lattice is viewed as having a two-site basis per unit cell of a tight-binding model. The two sites correspond respectively to the projection of the vertices and the projection of the bonds of the triangular lattices onto the horizontal axis and are depicted by the ticks and the links respectively in~\Fig{fig:dimer}. The sites corresponding to the projected vertices will be taken to be even sites (ticks in~\Fig{fig:dimer}) and the neighbouring sites coming from the projected bonds (links in~\Fig{fig:dimer}) will be denoted as odd sites. 

We place hardcore bosons on this chain with the following convention: a downward slopping segment of the path in the triangular lattice corresponds to placing a boson on the odd site (links in~\Fig{fig:dimer}) and an upward segment of the path corresponds to an empty odd site. A horizontal segment of the path in the triangular lattice corresponds to a placing boson on the even site (ticks in~\Fig{fig:dimer}), as depicted in \Fig{fig:dimer}. As in the Q6VM, we take one charge at the end of the electric field line to be at the origin, while the other one is placed at a site located at ${\bf r}=(2 \ell_x, \ell_y)$ in the dimer lattice. It is often convenient to visualise triangular lattice as a square lattice rotated by $45^\circ$, this makes the path resemble the one in the Q6VM except that now it is also allowed to occupy one of the diagonals of the square lattice. In this picture one charge would be placed at ${\bf r'}=(\ell_x, \ell_y)$. In the $1d$ representation, the chain would have a total number of sites (counting both even and odd sites) given by $4\ell_x-1$ and a total number of bosons $N = \ell_x - \ell_y$. 

Although this convention assigns a unique boson configuration to every allowed path between the two charges, not every boson configuration corresponds to a physical path. Following the rules described above, there is no allowed dimer configuration corresponding to two bosons occupying two nearest neighbor sites, one site being even and the other being site odd. Additionally, there are no allowed configurations with two bosons occupying adjacent second nearest neighbor sites which are both even; namely if a boson sits at site $2i$ there cannot be another boson at site $2i+2$. These boson configurations need to be projected out. This is easily accomplished because the two constraints are local and can be enforced by adding a large energy penalty to those configurations with a term of the form:
\bea
\label{eq:con}
H_{con} &= U \sum_i n_i n_{i+1} + U \sum_{i \, even} n_i n_{i+2}, \quad U \to \infty \ .
\eea
After these two constraints are imposed, the path and boson Hilbert spaces are in one to one correspondence and one can show that the filling of the model is $\nu \sim \frac{1}{4}(1- \ell_y/\ell_x)$. Notice that the constraints impose that the maximum allowed filling is $\nu = 1/2$.

Let us now describe the quantum dynamics induced by the action of the RK Hamiltonian~\eqref{eq:ham} on the dimer configurations. There is a single kinetic term that flips plaquettes and acts on the path in the triangular lattice by flipping a horizontal segment into either up-down or down-up segment as depicted in \Fig{fig:dimer}. Therefore, just as in the Q6VM, this term acts on the bosons simply as a single particle-hopping term given by
\bea
H_{hop} &= - \sum_{i} t b^\dag_i b_{i+1} + h.c. \ . \\
\eea
Now we describe the potential terms which are diagonal in the path configuration and in the boson occupation basis. When the bosons lie in the odd sites (namely when the path has no horizontal segment) then a flippable plaquette appears at every corner of the path, which corresponds to every link of the 1D lattice in which there is a change of occupation of the odd sites, in complete analogy to what was found in the Q6VM. This leads to a potential energy term for the odd sites of the form:
\bea
\label{eq:pot1}
H_{pot,1} = V \sum_{i \, odd} (n_{i} - n_{i+2})^2 \ .
\eea
Now, every horizontal segment of the path has two adjacent flippable plaquettes as depicted in \Fig{fig:dimer}, and therefore there is a term that acts as a the local chemical potential shift on the odd the sites of the form
\bea
\label{eq:pot2}
H_{pot,2} &= 2V \sum_{i \, even} n_i \ .
\eea
In addition when there is a horizontal segment of the path that is adjacent to a downward segment, there is one less flippable plaquette than those counted by the terms in \Eqs{eq:pot1}{eq:pot2} and therefore such energy is removed by a term of the form:
\bea
H_{pot, 3} &= - V \sum_{i } n_i n_{i+3} \ .
\eea
Thus the complete Hamiltonian is a sum of the terms above; explicitly
\bea
\label{eq:dimer}
H_{dimer} &= H_{hop} + H_{pot,1}+H_{pot,2}+H_{pot, 3}+H_{con}.
\eea
This Hamiltonian generalizes the one first obtained by Orland in \Ref{Orland1994} to non-zero values of $v$. Notably, unlike the Q6VM, this model does not reduce to free fermions at $V=0$ due to the presence of the constraint interactions in \Eq{eq:con}. For periodic boundary conditions the Hamiltonian is invariant under translations by two:
\bea
\label{eq:T}
\null [H, T] = 0, \quad T b^\dag_i T^{-1} = b^\dag_{i+2}
\eea
In addition to this symmetry, there is a particle-hole-like symmetry, which can be understood in the triangular lattice picture as the symmetry that reflects the path along the horizontal axis that intersects the charge at the origin. Viewing the triangular lattice as a square lattice rotated by 45$^\circ$ makes this symmetry look analogous to the particle-hole symmetry we encountered in the Q6VM. This symmetry is presumably responsible for a degeneracy we observe in numerics at quarter filling ($\nu=1/4$), between momenta at $k$ and $k+\pi$ for anti-periodic boundary conditions but we have not found a rigorous proof for this.

\begin{figure}

\includegraphics[scale=1]{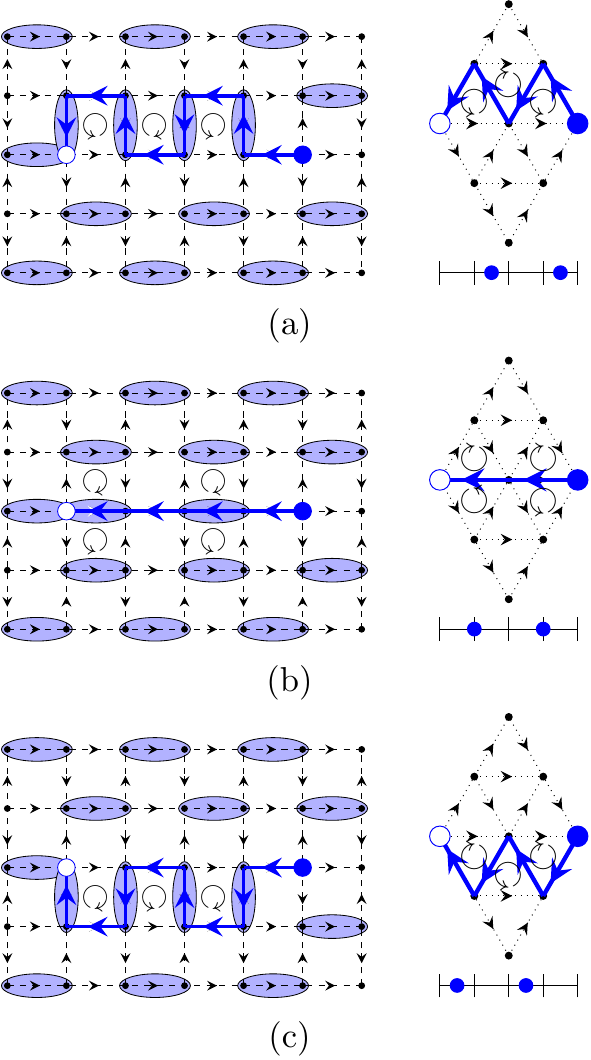}
\caption{Odd and even CDW states. Figures (a) and (c) show CDW on odd sites and are related under the particl-hole-like symmetry, while (b) shows the CDW on even sites. }
\label{fig:CDW}
\end{figure}

\subsection{Phase diagram}

In the remainder we restrict to the case of quarter filling where the charges are only horizontally displaced from each other by $\ell_x = L, \ell_y = 0$ on the triangular lattice, and corresponds to the largest Hilbert space which is expected to be closest precursor to the behavior of the full two-dimensional underlying QDM. We consider a periodic chain of length $\ell = 4 L$ in order to exploit the conservation of total many-body momentum of the $L$ bosons. We begin considering the limit $V \to - \infty$, in which the plaquette resonance terms acts as a perturbation on classical ground states, which in the boson language can be thought of as charge density wave (CDW) states with a definite occupation of the sites. Remarkably, and unlike the Q6VM case, there are two distinct degenerate CDW states in this limit for the QDM. They correspond to the bosons occupying either the even or the odd sites. Each of these states spontaneously breaks the translation symmetry defined in \Eq{eq:T}, and each of them has a symmetry related copy leading to a total of four ground states in the thermodynamic limit, depicted in Fig.~\ref{fig:CDW}. The explicit wavefunction in the $V\to -\infty$ limit of the two distinct states is:
\bea
\ket{odd} = \prod_{i=0}^{L-1} b^\dag_{4i + 1} \ket{0}, \quad \ket{even} = \prod_{i=1}^L b^\dag_{4i -2} \ket{0}\\
\eea
and their symmetry related copies are obtained by translating those above by two units. In the path picture on the triangular lattice, they are either maximally jagged (odd) or maximally straight (even), each having the maximal number (2$\ell$) of flippable plaquettes (see Fig. \ref{fig:CDW}). 

\begin{figure}
\begin{center}
\includegraphics[width=8cm]{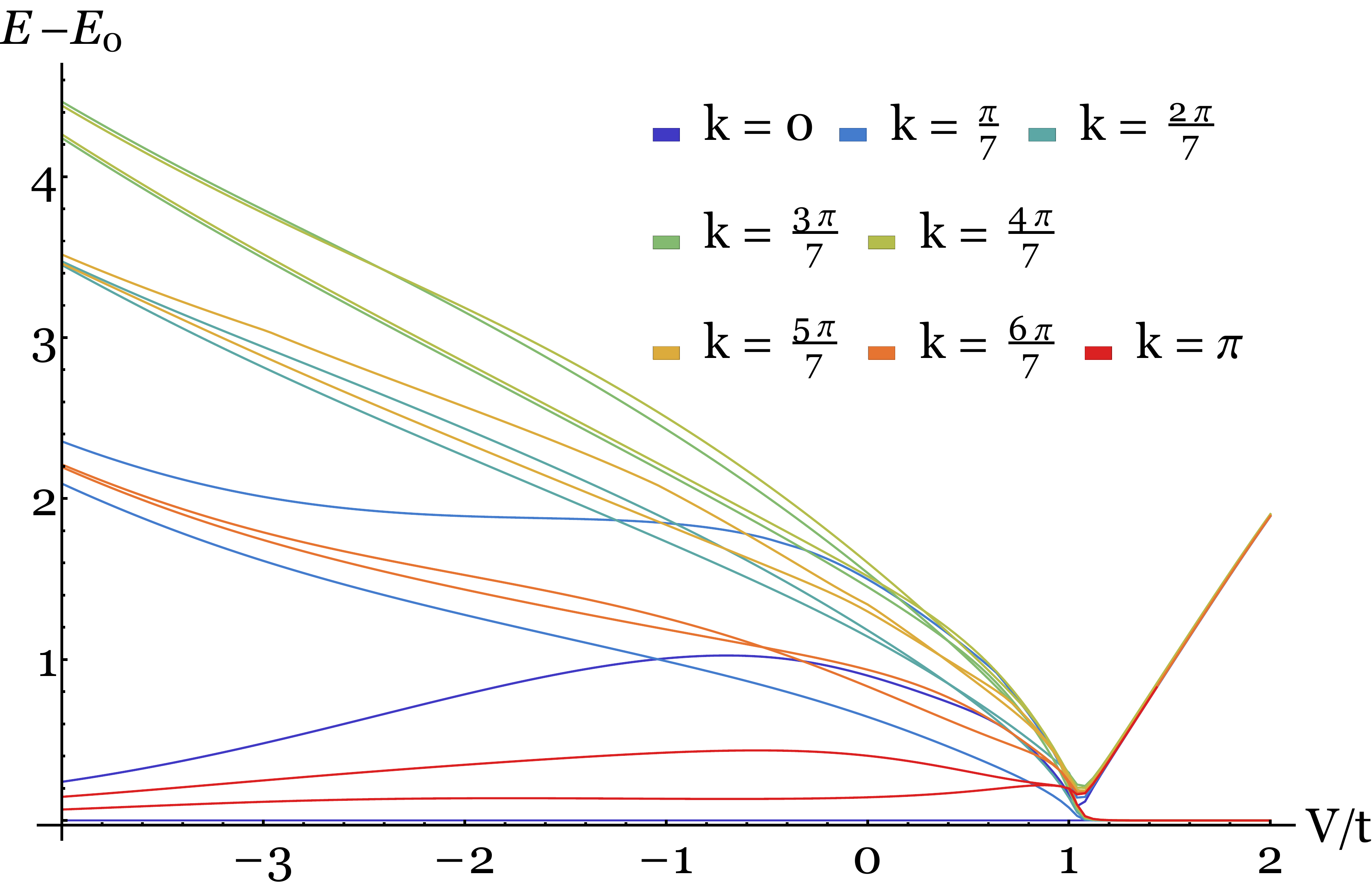}
\caption{The two lowest energy states in each $k$ sector for $L=7$ are shown relative to the ground state energy at $\phi =0$. At large negative $v$, the odd and even charge density waves at $k=0, \pi$ become degenerate. The discontinuity in the first derivative at the RK is visible. On the right, the lowest energy state at each $k$ become approximately degenerate, forming the phase-separated eigenstates in \Eq{eq:phasesep}.}
\label{fig:dimerspec0}
\end{center}
\end{figure}

Because there is no symmetry relating these states, quantum fluctuations will select a unique true ground state at finite $t$ out of these two. As we will see, however, their competition is delicate and is only resolved at fourth order in perturbation theory in $t$. All odd terms in $t$ must vanish, because the sign of $t$ can be changed by a gauge transformation, and thus the perturbation series begins at $t^2$. These are easily seen to be degenerate because in both CDWs the particle can hop one site in either direction. At fourth order, this degeneracy splits due to competition between single-particle and two-particle processes. On the odd sublattice, a single particle is able to hop two spaces in either direction and return to its original site, thus lowering the energy by delocalizing. In contrast, $H_{con}$ prevents this on the even sublattice. However, the pair encounters raise the energy for $V <0$. On the odd sublattice where particles are more mobile, these interactions raise the energy. Explicitly, we find
\bea
\frac{E_{odd}}{L} &= 2V - \frac{2t^2}{|V|} + \frac{t^4}{|V|^3} + O(t^6) \\
\frac{E_{even}}{L} &= 2V - \frac{2t^2}{|V|}  + 0 + O(t^6)\\
\eea
leaving the even sublattice CDW as the true ground state. The spectrum at $L=7$ is shown in \Fig{fig:dimerspec0}; the 4 ground states at $k=0, \pi$ are evident, along with the excitations above them. 

As we raise $V$ from $-\infty$, the bosons spread into the other sublattice in order to lower their energy by delocalizing under the increasingly strong kinetic term. This allows for the possibility of a phase transition like the one we encountered at $v=-1$ in the Q6VM. Even for small $v$, the model is strongly interacting due to $H_{con}$, and it is non-trivial to study the system here analytically. From the original 2D Hamiltonian, and as will be evident from the numerics, the system has a phase transition at the RK point $v = 1$ above which we can perform a strong coupling expansion. For $v>1$, particles in the odd sublattice experience the nearest neighbor attraction resulting from \Eq{eq:pot1}, and the constraints impose no restriction for them to cluster on the odd sublattice. This state is phase separated and has vanishing energy density, much like the ferromagnetic phase of the XXZ chain. This is most easily seen on the triangular lattice where the classical $V \to \infty$ ground state is given by a large triangular path, degenerate with its reflection under particle-hole symmetry. For periodic boundary conditions, there would be two flippable plaquettes, the second appearing where the string reconnects. Because of translational symmetry, a larger quasi-degenerate manifold would appear due to the delocalisation of the domain walls. At quarter filling, these $2L$ quasi-degenerate states are given by
\bea
\label{eq:phasesep}
\ket{k} &= \frac{1}{\sqrt{2L}} \sum_{n=0}^{2L-1} e^{-ik n} \prod_{i=1}^L b^\dag_{2i+1 +2n} \ket{0}, \quad k \in \frac{2\pi}{2L} \mathbb{Z} \ , \\
\eea
and are all degenerate as $V \to \infty$. One can perform standard perturbation theory to find this energy to be:
\bea
E &= 2V - \frac{2 t^2}{V} +  \dots \ .
\eea
This is twice the domain wall energy (twice because of the periodic boundary conditions). Since the uniform ground state has exactly zero energy density, there is strictly zero string tension, and all the energy is localized within the domain wall. This encodes the same kind of subdimensional deconfinement that we previously encountered in the case of the Q6VM between the monomer charges at the end of the electric field line.

\begin{figure}
\begin{center}
\includegraphics[width=8.5cm]{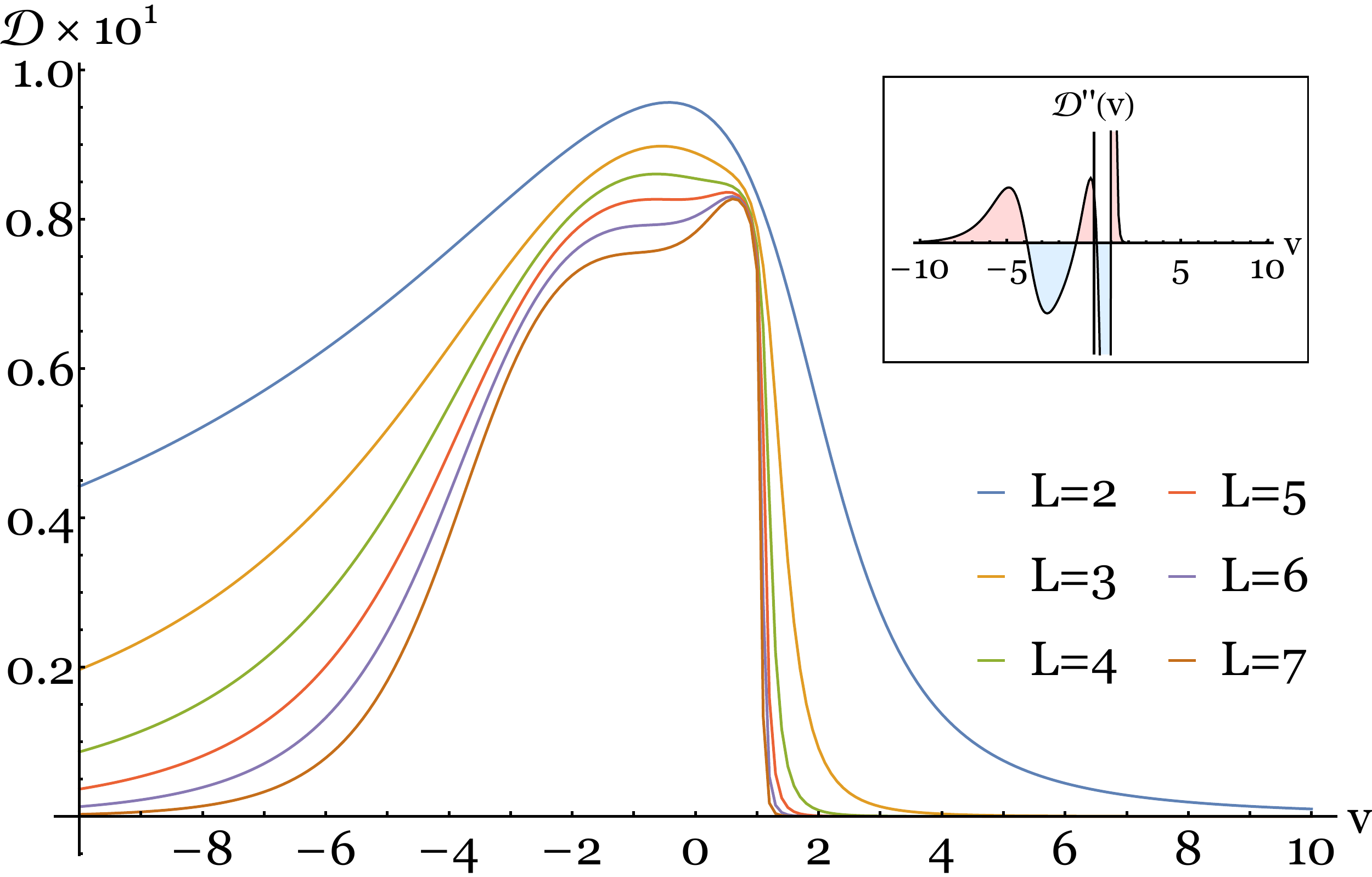}
\caption{Drude weight of $H_{dimer}$ is computed for systems $L=2, \dots, 7$ via exact  diagonalization for the ground state in the $k=0$ momentum sector. The RK point appears sharply at $v=1$ and there is evidence for an intermediate conducting phase. Unlike the Q6VM, there is much more structure in the putative critical region; as shown in the inset, the second derivative of the Drude weight (for $L = 7$) changes sign three times.}
\label{fig:dimerphase}
\end{center}
\end{figure}

As we have discussed, the two limits of $V \to \pm \infty$ lead to classical insulating CDW-type states. Metallic Luttinger-liquid-like states could appear in between these two limits. To explore this possibility quantitatively, we calculate the Drude weight from the exact diagonalization of the system for $\ell=2, \dots, 7$ with $\phi$ flux through the periodic system. In analogy with the Q6VM case, the twisted boundary conditions are implemented by adding one extra site at the end and localizing the phase change $\phi$ in final bond. We implement the numerical code using only the physical states derived from the string picture to avoid the extended Hilbert space of the bosons. The Drude weight at a selection of system sizes is shown in \Fig{fig:dimerphase}. 

Our numerical results confirm the intuition above --- that there appears a metallic region intervening between the two insulating phases realized at $V \to \pm \infty$, which we interpret as the natural 1D precursor to the 2D plaquette phase. The RK point appears clearly. There is evidence in the softer falloff at $V<0$ for a second critical point like in the Q6VM with a conducting region in between. We caution that our numerics are restricted to small system sizes, preventing us from making systematic extrapolations to the thermodynamic limit. Interestingly, the behavior of the Drude weight as a function of $v$ is notably different from the Q6VM result. In the Q6VM we see that the second derivative of the Drude weight with respect to $v$ changes sign only two times as function of $v$ near the critical points between the insulating and metallic phase, as shown in the inset of~\Fig{fig:XXZ}. However, the second derivative of the Drude weight appears to change sign ${\it four}$ times in the QDM, as shown in the inset to~\Fig{fig:dimerphase}. This could indicate the presence of more than one phase intervening between the two classical ground states at $V \to \pm \infty$, but verifying such speculation would certainly require accessing much larger system sizes, which should be possible in DMRG studies. 

\section{Summary and Discussion}

We have studied the problem of a single fluctuating quantum electric field lines connecting two isolated charge monomers in the QDM and Q6VM. By constructing these isolated strings on top of trivial inert vacua that appear as ground states to the right of the RK point, we have been able to map the problems exactly onto conventional one-dimensional systems with local Hamiltonians. 

In the case of the Q6VM model, the electric field line maps exactly onto the spin-$1/2$ quantum XXZ chain, which is an exactly solvable and well understood. This reveals three distinct phases which correspond to the Ising ferromagnet, the XY magnet, and the Ising antiferromagnet. These three phases are natural one-dimensional precursors to the full two-dimensional phases, seen in numerical studies of the Q6VM~\cite{Shannon,Banerjee2013}. In particular the resonant plaquette phase appears as the Luttinger liquid-like XY magnet state. This mapping has allowed us to reveal an underlying SU(2) symmetry in the RK point for large winding sectors, which corresponds to the critical point between the Ising and the XY magnet in the XXZ chain. This SU(2) symmetry leads to a {\it perfect} charge deconfinement in which the string tension separating charges vanishes even  for finite string length. It also allows to understand the sub-dimensional deconfinement present in the staggered phase~\cite{Shannon} as a form of phase separation and domain wall formation that occurs in the Ising magnet when the global magnetization is fixed. 

In the case of the QDM model, the electric field line maps exactly onto a one-dimensional lattice of hard-core bosons with two sites per unit cell (or equivalently a two-leg spin-$1/2$ ladder). We have not been able to solve this problem analytically but have been able to understand perturbatively the phases that occur far away from the RK point. These two phases correspond to a CDW state (analogous to the Ising AFM in the Q6VM case) and a phase separated state (analogous to the phase separated Ising FM in the Q6VM case). These two states are the natural precursors to the columnar and the staggered phases of the full two-dimensional QDM. Subdimensional deconfinement appears in the phase separated state, corresponding to the deconfinement in the staggered dimer phase~\cite{Batista}. By numerically computing the Drude weight, we have found evidence for a liquid state intervening between the two crystals that exist away from the RK point. We interpret this liquid, delocalized phase as the one-dimensional precursor of the plaquette valence bond solid in the full two-dimensional problem.

The resemblance of the quasi-one-dimensional single electric field line problems that we have studied and the full two-dimensional problems indicates that the behavior of the latter might be understood by thinking of them as a closely packed array of electric field lines which by themselves are undergoing non-trivial phase transitions. More specifically, as we described in \Sec{sec:structure}, the decoupling of the Hilbert space into winding sectors can be interpreted as a conservation law for the the total number of electric field lines. The zero winding sector where the global ground state of the full Hilbert space resides at the left of the RK point contains a large number of such electric field lines. These lines can be viewed as bosonic strings with hard-core interactions so that the electric field lines do not overlap. This perspective provides a natural understanding of why the crystalline phases of the one-dimensional electric field line survive in the fully two-dimensional multistring case. However, these strong interactions are presumably responsible for the freezing out of the quantum fluctuations of the Luttinger liquid type phase that we have encountered, transforming it into the resonant plaquette crystal state that is seen in numerical studies of the full two-dimensional problem of the six-vertex model. We hope that in the future, more systematic numerical studies of our current setting and of its generalizations to the few strings problems might offer an alternative window the behavior of the less well understood aspects of the presumed resonant plaquette phase of quantum dimer model.

\section{Acknowledgements}

We are thankful for valuable discussions with Cristian Batista, Roderich Moessner, Alet Fabien, Gregoire Misguich, Yi-Ping Huang, Markus Heyl, and Arnab Sen. Special thanks to Debasish Banerjee for teaching us many aspects of these models and of lattice gauge theory.

\bibliography{dimers-arxiv}
\bibliographystyle{aipnum4-1}

\setcitestyle{numbers,square}
\end{document}